\documentstyle[12pt]{article}
\def\pp{\hfill\par \vspace{\baselineskip}}

\begin{document}

\input epsf

\newcommand{\oeawpreprint}[2]
{
\noindent
\begin{minipage}[t]{\textwidth}
\begin{center}
\framebox[\textwidth]{$\rule[6mm]{0mm}{0mm}$ 
\raisebox{1.3mm}{Institut f\"ur Hochenergiephysik der \"Osterreichischen
Akademie der Wissenschaften}}

\vspace{2mm}    \rule{\textwidth}{0.2mm}\\
\vspace{-4mm}   \rule{\textwidth}{1pt}
\mbox{ }    #1    \hfill    #2   \mbox{ }\\
\vspace{-2mm}   \rule{\textwidth}{1pt}\\
\vspace{-4.2mm} \rule{\textwidth}{0.2mm}
\end{center}
\end{minipage}

}   

\oeawpreprint{November 1995}{HEPHY-PUB 634/95}

\smallskip
\mbox{ }\hfill hep-ph/9609288

\vspace*{20mm}

\begin{center}
{\Large\bf Pion interferometry with higher-order cumulants\footnote{
To be published in the Proceedings of the 
XXV-th International Symposium on
Multiparticle Dynamics, Star\'a Lesn\'a, Slovakia,
12--16 September 1995}
}\\

\mbox{ }\\
H.C.\ Eggers\footnote{
              Present address: Department of Physics, University of
              Stellenbosch, 7600 Stellenbosch, South Africa;
              E-mail: eggers@bohr.sun.ac.za},
B.\ Buschbeck and P.\ Lipa \\
\mbox{ }\\
{\it 
     Institut f\"ur Hochenergiephysik der \"Osterreichischen
     Akademie der Wissenschaften,} \\
{\it 
     Nikolsdorfergasse 18, A--1050 Vienna, Austria}\\
\end{center}

\vspace*{0.9cm}

\begin{abstract}
We have measured second- and third-order cumulants in UA1 data ($\bar
pp$ collisions at 630 GeV/c). Rather than quoting numerical values for
source parameters, we have used these in three checks to test the
``quantum statistics'' theory for consistency over these cumulants. In
the process, we have found a method for folding theoretical
correlation functions with experimental one-particle distributions.
Our preliminary results appear to indicate that, for the specific tests
performed, the data contradicts the theory.
\end{abstract}

\vspace*{1.4cm}


\section{Introduction}

Pion interferometry has been a part of particle physics
for several decades \cite{Zaj88a}. The main sub-branch of this science, 
concerning
itself with Bose-Einstein correlations, endeavours to elicit
information on the size, shape and temporal evolution of
the source emitting pions. This is based on an analogy
between optical intensity interferometry and quantum mechanical
interference between incoherent pion amplitudes.
\pp

Measurements of correlations between identical particles contain,
besides amplitude interference, a plethora of other effects such as
coherence, decaying resonances, variations in impact parameter and
momentum distribution, contamination by kaons and protons, 
final-state interactions etc.  Understandably, a solid basis for
subtracting all such effects has been singularly hard to create. Given
the number and degree of theoretical and experimental uncertainties
entering even second-order correlations, the corresponding higher-order
measurements have received only scant attention:  if it is hard to
extract source parameters in an honest and unambiguous way from
second-order correlation data, it probably becomes even harder for
third order.
\pp

In the present paper, we elect to take a different approach:
We measure higher-order correlations not so much with a view
to extracting source parameters or ``true'' Bose-Einstein
correlations, but in order to perform {\it consistency checks}.
While few theorists have so far worked out the implications
of their respective models for higher orders, this is in
principle possible, and some examples of higher-order
predictions exist \cite{Wei89a,Biy90a,Lyu91a,Plu92a,And93a}.
If a given theory provides formulae
for both second and higher orders, then these should apply to
the corresponding data using the same parameter values
throughout.
\pp

With the aid of rapidly-improving measurement technology, we are
attempting to put such predictions to the test. While the results
reported here are quite preliminary in nature, they hopefully point the
way to more general and sophisticated testing of theories rather than
just measuring their respective parameters.  {\it Falsifying\/}
theories is arguably the best (some would say the only) way of making
progress in a confused situation \cite{Pop35a}.
\pp

Our tools for performing these consistency checks are {\it cumulants\/}
and the {\it correlation integral\/} \cite{Lip92a,Egg93d,Egg93a}.
Cumulants, in subtracting out trivial lower-order contributions, have
proven far more sensitive than the corresponding moments; their
implementation in various forms of the correlation integral has, at the
same time, improved statistical accuracy to a degree where such
measurements have become meaningful.

\section{Quantum statistics theory}

The test we shall be reporting here is confined to
one particular variable, the four-momentum difference 
$q_{ij} = [(\vec p_i - \vec p_j)^2 - (E_i - E_j)^2]^{1/2}$.
For this variable, the second and third-order 
cumulants are \cite{Egg93d,Ber77a}
\begin{eqnarray}
\label{cua}
C_2(q) &=& \rho_2(q) - \rho_1{\otimes}\rho_1(q) \,,  \\
C_3 
&=& \rho_3 - \sum_{(3)} \rho_2{\otimes}\rho_1
    + 2 \rho_1{\otimes}\rho_1{\otimes}\rho_1  \,,
\end{eqnarray}
where the third order quantities are functions of the three pair
variables $(q_{12},q_{23},q_{31})$. These cumulants, including the
crossed ``${\otimes}$'' quantities and event-mixing normalizations
can be found from data samples in a precisely prescribed
algorithm \cite{Egg93d}.
\pp

The quantum statistics (QS) theory itself has a long and distinguished 
tradition \cite{Fow78a,Gyu79a};
the version we concentrate on is based on analogies
to quantum optics (for details, we refer the reader to
Refs.\ \cite{Biy90a,Plu92a,And93a}).
Briefly, the main features of interest to us are:


\noindent  
{\bf a)}
The pion field is split up into a ``coherent'' and a
``chaotic'' part:
\begin{equation}
\label{cub}
\Pi(x) = \Pi_0(x) + \Pi_{ch}(x) \,.
\end{equation}
\noindent  
{\bf b)}
The ratio of chaotically created pions to the total 
number of pions is embodied in the ``chaoticity parameter'', 
\begin{equation}
\label{cuc}
p = \langle n_{ch} \rangle / \langle n_{ch} + n_0 \rangle \,.
\end{equation}
\noindent  
{\bf c)}
Much of the dynamics is contained within the 
normalized field correlator,
\begin{equation}
\label{cud}
d_{ij} \equiv  
{\langle \Pi_{ch}^{\dag} (\vec k_i)  \Pi_{ch}(\vec k_j) \rangle
\over
\left[ \,
  \langle \Pi_{ch}^{\dag} (\vec k_i) \Pi_{ch}(\vec k_i) \rangle \;
  \langle \Pi_{ch}^{\dag} (\vec k_j) \Pi_{ch}(\vec k_j) \rangle \,
\right]^{1/2} }   \ \;,
\end{equation}
(where $\langle A \rangle = {\rm Tr}(\rho A)$ is the
ensemble average over states, weighted by the density matrix
$\rho$)
which is closely related to the Fourier transform of the chaotic
field source functions. 


\noindent  
{\bf d)}
Working out two-point, three-point and higher-order averages, 
this theory of quantum statistics predicts unambiguously
the normalized moments and cumulants of all orders. When relative
phases are neglected, the first three ``QS cumulants'' 
of interest are \cite{And93a}
\begin{eqnarray}
\label{cue}
k_2 \equiv {C_2 \over \rho_1{\otimes}\rho_1}
   &=& 2p(1-p)d_{12} + p^2 d_{12}^2 \,, \\
\label{cuf}
k_3 \equiv {C_3 \over \rho_1{\otimes}\rho_1{\otimes}\rho_1}
   &=& 2p^2(1-p)[ d_{12}d_{23} + d_{23}d_{31} + d_{31}d_{12} ]
      + 2p^3 d_{12} d_{23} d_{31} \,,\\
\label{cug}
k_4 \equiv {C_4 \over \rho_1{\otimes}\rho_1{\otimes}\rho_1{\otimes}\rho_1}
&=& \sum_{(24)} p^3(1-p) d_{12} d_{23} d_{34} 
\nonumber\\
&&\ {+} 2p^4 [d_{12}d_{23}d_{34}d_{41} + 
              d_{12}d_{24}d_{43}d_{31} + 
              d_{14}d_{42}d_{23}d_{31} ]\,,
\end{eqnarray}
where the brackets under the sum indicate the number of permutations.
These cumulants are functions of 1, 3 and 6 pair variables 
$q_{ij}$ respectively.
Note the combination of ``ring''-- and ``snake''--like structures in
the combinatorics.
\pp

While in principle calculable from a given density matrix, the
correlator is usually parametrized in a plausible and/or convenient
way. Specifically, the parametrizations we shall be testing are, in
terms of the 4-momentum difference correlators $d_{ij} {=} d(q_{ij})$,
\begin{eqnarray}
\label{cuh}
{\rm gaussian{:}\ \ \ \ \ \ }    
d_{ij} &=& \exp(-r^2 q_{ij}^2) \,, 
\mbox{\ \ \ \ \ \ }
\\
\label{cui}
{\rm exponential{:}\ \ \ \ \ \ } 
d_{ij} &=& \exp(-r q_{ij})     \,, 
\mbox{\ \ \ \ \ \ }
\\
\label{cuj}
{\rm power\ law{:}\ \ \ \ \ \ }   
d_{ij} &=& q_{ij}^{-\alpha}  \,.  
\mbox{\ \ \ \ \ \ }
\end{eqnarray}

\section{UA1 data}

We have measured second- and third-order normalized cumulants using a
sample of about 160,000 minimum bias events taken with the UA1 detector
for $p\bar p$ collisions at 630 GeV/c.  For details of the detector
and other experimental information regarding particle pairs,
the reader is referred to Ref.~\cite{UA1-93a} The following cuts were
applied to this sample:  $-3 \leq \eta \leq 3$, $p_\perp \geq 0.15$
GeV, $45^\circ \leq \phi \leq 135^\circ$ (by means of this ``good
azimuth'' cut, our statistics were reduced considerably but acceptance
corrections due to the ``holes'' in the UA1 detector at small $\phi$
were thereby avoided). Cumulants were calculated for positives and
negatives separately and then averaged to yield ``like-sign'' values.
No Coulomb corrections were applied.

\section{Fits to the second order cumulant}

In Figure~1, we show the second-order like-sign differential
cumulant $\Delta K_2 = (\int \rho_2 / \int \rho_1{\otimes}\rho_1 )$
$-1$, where numerator and denominator are integrals over bins
spaced logarithmically between $q = 1$ GeV and 20 MeV\footnote{
It should be remarked that previous work has shown the utility
of using logarithmic rather than linear binning: much of what is
interesting in correlations happens at small $q$, and
this region is probed better by using logarithmic bins.}.
Fits to the data were performed using the three parametrizations
(\ref{cuh})--(\ref{cuj}), either in the full QS form (\ref{cue}) or in
a simple form $\Delta K_2 = p d_{12}$. All fits shown
include, besides the free parameters $p$ and $r$ (or $\alpha$), an
additive constant as free parameter. These additive constants,
necessary because UA1 data is non-poissonian in nature, will be
commented on further below. Best fit parameter values obtained were $p
= 0.66 \pm 0.07$, $r = 1.16 \pm 0.05$ fm for the QS exponential and $p
= 0.05 \pm 0.01$, $\alpha = 0.64 \pm 0.05$ for the QS power law.
Goodness-of-fits were $\chi^2/{\rm NDF} = 1.3,\ 4.2$ and 11.5 for QS
power, exponential and gaussian respectively.
\pp

To check its influence on fit values, the data point at smallest $q$,
being of doubtful quality, was excluded; the resulting fit values do
not differ much from the full fit.
We note that the QS exponential misses the last three points (apart
from the point at $q=20$ MeV) and that the power laws (single or QS)
appear to do the best job. The gaussian fits are too bad to warrant
further attention and will be neglected from here on.  Similar
conclusions were reached by UA1 earlier \cite{UA1-93a}.

\begin{figure}
\centerline{
\epsfysize=100mm
\epsfbox{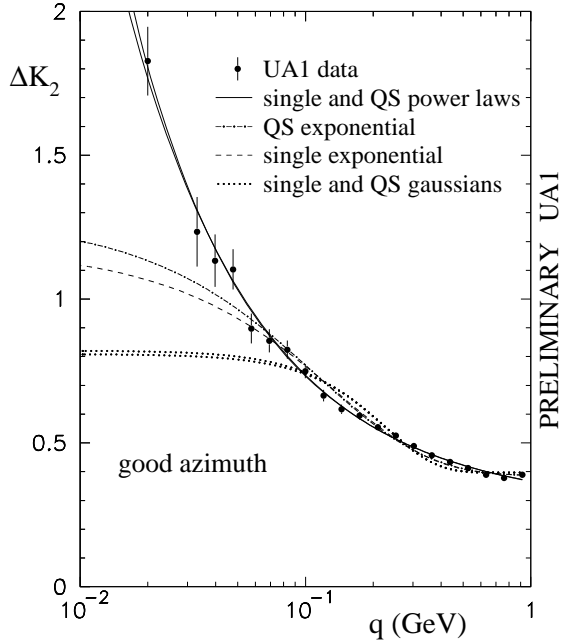}
}
\vspace*{-8mm}
\caption{Second order like-sign UA1 cumulant with fits using various
parametrizations for $d$ inside both the quantum statistical (QS)
formula (\ref{cue}) and a simple parametrization $\Delta K_2 = pd$. }
\label{fig1}
\end{figure}

\section{Consistency checks with third order cumulants}

As stressed already, we are interested not so much in obtaining
numerical values for parameters but rather in using these to check the
theoretical formulae (\ref{cue})--(\ref{cuj}) for consistency with the
data. Three separate checks were performed: two based on
approximations, the third involving a novel approach tentatively called
``theory${\otimes}$experiment'' which will be explained in Section 5.2.
\pp

\subsection{Approximate checks}

Third-order correlations and cumulants are
functions of the three pair variables $(q_{12},q_{23},q_{31})$,
so that the question arises how best to view such three-dimensional
correlations. The easiest projection involves setting
the pair variables equal \cite{Biy90a,Plu92a,UA1-92a},
$q_{12}{=}q_{23}{=}q_{31}{\equiv}q$,
so that Eq.\ (\ref{cuf}) reduces to the simple formula
\begin{equation}
\label{cuk}
k_3(q) = 6p^2(1-p)d^2 + 2p^3d^3 \,. 
\end{equation}
Experimentally, however, the prescription of {\it three\/} mutually
equal $q$'s is so restrictive as to make measurement impossible. The
usual way out \cite{UA1-92a,NA22-95a} has been to include all triplets whose 
{\it mean\/}  of the three $q_{ij}$'s is equal to a given $q$ while still
applying Eq.\ (\ref{cuk}) (the effect of this approximation has, to our
knowledge, not been checked).
\pp

The second approximation involves setting $p \equiv 1$ without
restricting the pair variables. Fortuitously, Eqs.~(\ref{cuf}),
(\ref{cui}) then become $k_3 = 2\exp[-r(q_{12}+q_{23}+q_{31})]$, so
that a simple change to the ``GHP~sum'' variable $S =
(q_{12}{+}q_{23}{+}q_{31})$ does the trick.
\pp

In Figure~2, we show the UA1 third-order cumulant $\Delta K_3$
as a function of the GHP~sum variable $S$. 
The lower line represents the first approximation,
i.e.\ formula (\ref{cuk}) using the exponential parametrization
(\ref{cui}) and best-fit values from
$\Delta K_2$ plus an arbitrary additive constant.
(Similar approximations using the gaussian form (\ref{cuh})
with the variable\footnote{
Note that the linear sum $S$ is quite distinct from the
pythagorean sum variable $Q$.} 
$Q \equiv \sqrt{q_{12}^2 + q_{23}^2 + q_{31}^2}$ 
and equal pair $q$'s have been used before \cite{NA22-95a}.)
The upper line, representing the second approximate check, was
calculated by first fitting $\Delta K_2$ with $p{=}1$ and an QS
exponential for $d$ to obtain $r = 0.89 \pm 0.02$ fm (not shown) and
then importing this value into $k_3(p{=}1) = 2\exp({-rS})$.
\pp

We see that, in both cases, the theoretical curves lie
well below the $\Delta K_3$ data. Even an arbitrary shift
by an additive constant does not improve the match because
of the different shape of the curves as compared to the data
points.

\begin{figure}
\centerline{
\epsfysize=90mm
\epsfbox{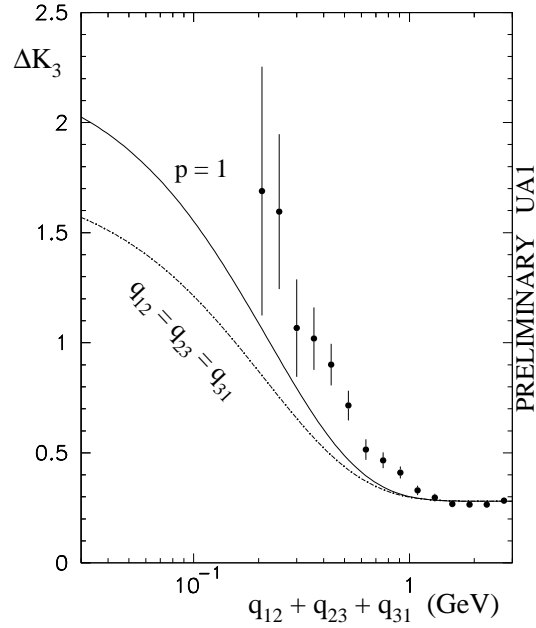}
}
\caption{Approximate test predictions, compared to the third-order UA1
cumulant using GHP sum topology.}
\label{fig2}
\end{figure}

\begin{figure}
\centerline{
\epsfxsize=135mm
\epsfbox{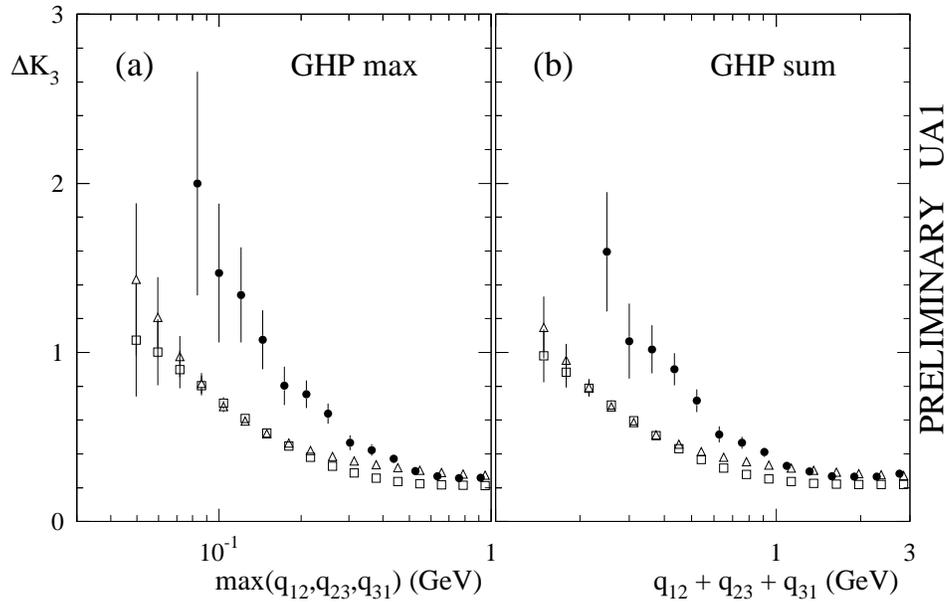}
}
\caption{
Third-order GHP max and GHP sum cumulants, together with
theory${\otimes}$ex\-pe\-ri\-ment predictions from QS theory and
parameter values from $\Delta K_2$. Filled circles represent UA1
data, triangles are predictions based on the QS power-law
parametrization; squares are QS exponential predictions.
}
\label{fig3}
\end{figure}

\subsection{The ``theory${\otimes}$experiment'' method}

The approximate consistency checks performed above are 
unsatisfactory for two reasons: first, because they rely on simplifications
of the formulae which may be unwarranted, second, because 
they are suitable only for the exponential parametrization
(or, using $Q$, for the gaussian equivalent). As shown above,
however, the data for $\Delta K_2$, while not excluding an
exponential form, would seem to prefer the power law ---
and the power law cannot be handled by these approximations.
A seemingly better methodology emerges, surprisingly, from
some considerations about normalization. 
\pp

{\it Theory\/} and theorists usually work with infinitesimally
differential normalized quantities; for example, the second-order
normalized cumulant is often written down as
\begin{equation}
\label{cul}
R(\vec k_1,\vec k_2) = 
{ \rho_2(\vec k_1,\vec k_2) \over \rho_1(\vec k_1) \rho_1(\vec k_2) } -1 
=
{C_2(\vec k_1,\vec k_2) \over \rho_1(\vec k_1) \rho_1(\vec k_2) }
\end{equation}
which is (implicitly) 
fully differential in the momenta $\vec k_1,\vec k_2$. 
Similarly, the normalized theoretical cumulants
$k_i = C_i/\rho_1{\otimes}\cdots{\otimes}\rho_1$ used above
assume essentially perfect measurement accuracy and infinite
statistics.
\pp

{\it Experimentally\/}, one can never measure fully
differential quantities; rather, the numerator and denominator
are averaged over some bin of finite size $\Omega$ (however small)
before the ratio is taken; for example
\begin{equation}
\label{cum}
\Delta K_2(\Omega) = 
{\int_\Omega C_2(q)\, dq \over 
 \int_\Omega \rho_1{\otimes}\rho_1(q)\, dq} \,,
\end{equation}
which approaches the theoretical cumulant $k_2(q)$
only in the limit $\Omega\to 0$. 
\pp

This observation can be converted into an exact prescription
for folding a given theoretical normalized quantity with
experimentally measured one-particle distributions. For
simplicity, we take second order quantities as an example.
Since trivially $C_2(q) = k_2(q) \, \rho_1{\otimes}\rho_1(q)$,
we can take $k_2$ from theory, $\rho_1{\otimes}\rho_1$ from
experiment and write exactly
\begin{equation}
\label{cun}
\Delta K_2(\Omega) = 
{\int_\Omega k_2^{\rm th}(q) \;
            \rho_1{\otimes}\rho_1^{\rm expt}(q) \, dq \over 
 \int_\Omega \rho_1{\otimes}\rho_1^{\rm expt}(q)\, dq} 
\equiv 
{\int_\Omega C_2^{{\rm th} {\otimes} {\rm expt}}(q) \; dq \over 
 \int_\Omega \rho_1{\otimes}\rho_1^{\rm expt}(q)\, dq}
\,.
\end{equation}
Correlation integral theory prescribes that \cite{Egg93d,Egg93a}
\begin{equation}
\label{cuo}
\rho_1{\otimes}\rho_1^{\rm expt}(q)
= \left\langle \left\langle \sum_{i,j} \delta[q - Q_{ij}^{ab}] 
  \right\rangle_{\!\!b} \right\rangle_{\!\!\!a}  ,
\end{equation}
where $Q_{ij}^{ab} 
= [({\vec {p_i}}^a - {\vec {p_j}}^b)^2 - (E_i^a - E_j^b)^2]^{1/2}$
is the four-momentum difference between two tracks $i$ and
$j$ taken from different events $a$ and $b$. 
Taking, for example, the QS cumulant (\ref{cue}) and the
exponential parametrization (\ref{cui}), this leads to
\begin{equation}
\label{cup}
C_2^{{\rm th} {\otimes} {\rm expt}} (q)
= \left\langle \left\langle \sum_{i,j} \delta[q - Q_{ij}^{ab}] 
  [ 2p(1-p) \exp(-rQ_{ij}^{ab}) + p^2 \exp(-2rQ_{ij}^{ab}) ]
  \right\rangle_{\!\!b} \right\rangle_{\!\!\!a}  ,
\end{equation}
which can be binned in $q$ or otherwise integrated.  In passing, we
observe that Eq.~(\ref{cun}) reduces to the theoretical $k_2$ for
infinitesimal $\Omega$ or for constant $\rho_1{\otimes}\rho_1$ as
required.
\pp

Clearly, this can be generalized to all possible moments and cumulants,
independently of variable or integration topology.  The procedure
exemplified by Eq.~(\ref{cup}) and its generalizations amounts to a
Monte Carlo integration of a {\it theoretical correlation function\/}
sampled according to the {\it experimental uncorrelated one-particle
distribution\/}; for this reason, we like to call it by the diminutive
``Monte Karli'' or ``MK'' for short.  MK can, of course, be implemented
only for fixed numerical values of the theo\-retical parameters, in this
case $p$ and $r$.  These must be determined either by more naive
fitting methods (and then checked for consistency) or by a very
cumbersome fitting procedure using the full event sample many times
over.
\pp

In Figure 3, the results of implementing the MK prescription
are shown. Besides the GHP sum topology used in (b),
we show in (a) a separate analysis using the ``GHP max''
topology \cite{Egg93a}, which bins triplets according to the largest
of the three variables, max$(q_{12},q_{23},q_{31})$.
Fit parameter values used for the respective power law
and exponential MK points were taken from the naive
QS fit to $\Delta K_2$ of Figure 1. (The consistency
of this procedure was checked by inserting these
parameter values back into the MK formulae for $\Delta K_2$
and finding agreement between UA1 data and MK predictions.)
Again, all MK points shown are determined only up to an additive
constant, so that the curves may be shifted up and down.
It is again clear, though, that the shape of third-order
cumulant data measured differs appreciably from that predicted
by the QS formulae and parameter values from $\Delta K_2$.
This conclusion holds independently of the topology used
and of the functional form taken for $d$.

\section{Discussion}

Concerning the fits to $\Delta K_2$, we have concluded that the
gaussian parametrization $d_{ij} = \exp(-r^2q_{ij}^2)$ is quite unsuitable,
while the exponential is better but not good. The best fit was obtained
using either a simple or QS (double) power law.  This confirms earlier
results \cite{UA1-93a}\cite{NA22-93a}. 
The fits were reasonably stable even when
excluding the point at smallest $q$, so that the effect is not due to
this last point.
\pp

Parameter values obtained from fits $\Delta K_2$ were then applied 
to third-order cumulant data in
three different checks. Both the two
approximations as well as the exact theory${\otimes}$experiment
(Monte Karli) prescription yielded predictions that did not match the data.  The
tests performed in this paper, namely checking three specific
parametrizations (gaussian, exponential and power-law) within one
specific variable $q$ for consistency between $\Delta K_2$ and
$\Delta K_3$ appear to indicate that, {\it under these specific
conditions}, the theory is contradicted by the data.
\pp

It should be clear, though, that this conclusion can at this stage be
preliminary and limited in scope only, for the following reasons:
\begin{itemize}

\item
The data shown is preliminary only and will have to
await further checks such as acceptance corrections, full-azimuth studies,
sensitivity to binning, etc.

\item
The most important caveat relates to the structure of the overall
multiplicity distribution.  The fact that UA1 data is not poissonian in
nature \cite{UA1-83a} can be seen immediately at large $q$ where
$\Delta K_2$ converges not to zero (as a poissonian cumulant would) but
to $\approx 0.4$. The same holds for $\Delta K_3$.  Theories, however,
are almost universally based on an overall poissonian: as can be easily
verified from Eqs.\ (\ref{cue})--(\ref{cug}), all cumulants tend to
zero for large $q$. The policy followed here, namely reconciling
poissonian theory with non-poissonian data by means of an additive
constant in the cumulants, is a sensible but hitherto poorly-understood
first step. The question of handling cumulants more adequately within a
non-poissonian overall multiplicity distribution is presently being
considered \cite{Liptb}.  We also hope that our results may goad
theorists into more careful consideration of their work with respect to
the implicit poissonian normalization used in most theories. See also
Ref.~\cite{And94a}.

\item
Closely related to these additive constants is the question of correct
normalization. Traditional lore in second order divides the density
correlation function (moment) $\Delta F_2$ by an additional
normalization factor $f$, taken as the moment at some large value of
$q$.  An alternative methods creates as many background pairs as
necessary to achieve the limit of unity for $\Delta F_2$. While the
third order moment can similarly be normalized to unity, the
prescription fails for third order cumulants. A brief scan of the
literature on third-order cumulants reveals that no adjustments were
made for possible non-poissonian multiplicity structure.
\cite{Biy90a,Lyu91a,Plu92a,And93a,Egg93d,Egg93a,Ber77a,UA1-92a,
Ken78a,DELPHI-95a,NA22-94a,Jur89a,Liu86a}

\item
Finally, one may mention possible changes to the present application of
the theory such as inclusion of relative phases in the correlators,
possible non-gaussian source currents, modelling the
momentum dependence \cite{And93a} of $p$, variable
trans\-for\-ma\-tions \cite{And93b} in $d$ and so on.

\end{itemize}

Beyond these caveats, the following points are of relevance to
the interpretation of our results:
\begin{itemize}

\item
No corrections for Coulomb repulsion \cite{Zaj88a}
were included. We could argue that the same Coulomb effects
that might shift $\Delta K_2$ data upward would increase $\Delta K_3$
data even more, since there are three pairs involved rather than one.
Even more convincing is the fact that $\Delta K_3$ data rises
more strongly than theoretical predictions even for large $q$
(several hundred MeV) where Coulomb repulsion is not expected to
be important. 

\item
Strictly speaking, the good power-law
fit to $\Delta K_2$ in itself is inconsistent:  QS theory
requires \cite{And93a} $\lim_{q\to 0}\; d(q) = 1$, while the power-law
parametrization diverges. Attempts to explain this in terms of variable
transformations \cite{And93b} or source size
distributions \cite{Gyu82a,Bia92a} may therefore provide useful starting
points in explaining the discrepancies in $\Delta K_3$.

\item
Track mismatching can lead to strong correlation effects because the
reconstruction program may split a single track into a closely
correlated pair. A great deal of effort in early experimental
intermittency studies went into creating clean ``split-track-finding''
algorithms \cite{Lipphd} and these are included in our analysis. We have
checked through additional small-$q$ cuts that mismatching does not
appear to explain our strong rises in the cumulants.

\item
The UA1 sample consists of $\sim$ 15\% kaons and protons which cannot
be distinguished from pions.  The effect that these would have on
$\Delta K_3$ is unclear.

\item
Resonances are known to increase $\Delta K_2$ at small $q$, the main
effect in {\it second order} deriving from interference between
``direct'' pions and resonance-decay products.\footnote{
We have also looked at like-sign correlations resulting from
resonance decay chains such as 
$(\eta^\prime \to \eta\pi^+\pi^-)$; $(\eta \to \pi^+\pi^-\pi^0)$:
using PYTHIA with the Bose-Einstein routines switched off, 
we find no significant like-sign correlations in PYTHIA from
resonances alone.} 
How and whether resonances would contribute to like-sign cumulants in
{\it third order} (and for values of $q$ of several hundred MeV shown
in $\Delta K_3$) is still quite mysterious.

\item
At this point, one could wonder whether it is wise to even attempt to
eliminate resonances from hadron-hadron collision data:  apart from the
theoretical and technical difficulties, what dynamical information does
the typical ``size''  $r < 1$~fm of a hypothetical ``source'' contain
that is more important than a cascade structure containing resonances
whose existence is beyond doubt? If the ``source'' is scarcely larger
than a nucleon, then how can one speak of incoherent or even classical
production of two pions?  And if one eliminates long-lived resonances,
then one would presumably still be left with the short-lived ones
rather than the holy grail of an abstract quantum mechanical ``source''.

\end{itemize}

\noindent
The results of the present paper may appear, at first sight, to
contradict the conclusion \cite{Plu92a}, based on an earlier UA1
paper \cite{UA1-92a},  that QS theory was compatible with
higher-order moments. The apparent discrepancy is explained by pointing
out that 1) measurement techniques have improved considerably since
then, 2) these techniques have permitted the present direct
measurements of cumulants, which are considerably more sensitive than
moments, and 3) even for these moment fits \cite{UA1-92a}, the radii
were not quite constant but showed a systematic increase.
\pp

Bose-Einstein correlation measurements with a view to extracting source
parameters are by now well-established in hadronic and heavy ion
phenomenology.  Our intention here was to show that consistency checks
between cumulants of different orders might be a second route to
learning something about the system: if by this method a given theory
can be tested already on a qualitative rather than quantitative basis,
then opportunities for feedback and improvement of such theories may
expand.

\bigskip

{\bf Acknowledgements:}
This work was supported in part by the Austrian Fonds zur F\"orderung
der wissenschaftlichen Forschung (FWF) by means of a Lise-Meitner
fellowship and an APART fellowship of the Austrian Academy of
Sciences.

\end{document}